\documentclass[aps,twocolumn,showpacs,floats,superscriptaddress]{revtex4}  
\usepackage{amsmath}
\usepackage{amssymb}
\usepackage{graphicx}

\newcommand{\lsim}{ \mathop{}_{\textstyle \sim}^{\textstyle <} }
\begin{document}

\preprint{ICRR-Report-518-2005-1}
\preprint{YITP-05-13}
\title{Investigation of Possible Dark Matter Direct Detection in Electron Accelerators}
\author{Junji Hisano,$^1$ Minoru Nagai}
\affiliation{ICRR, University of Tokyo, Kashiwa 277-8582, Japan }
\author{Mihoko M. Nojiri}
\affiliation{YITP, Kyoto University, Kyoto 606-8502, Japan}
\author{Masato Senami$^{1}$} 
\date{\today}

\begin{abstract}

  We investigate a possibility of neutralino dark matter (DM) direct
  detection in the future electron accelerators.  That is counting of
  high $p_T$ electron recoil events by neutralinos in halo.  If
  selectron and neutralino masses would be precisely measured in
  future collider experiments, the beam energy could be tuned so that
  the scatterings are dominated by on-pole selectron exchange. When
  selectron and neutralino mass difference is smaller than $O(10)$ GeV,
  the elastic cross section exceeds over micro barn. Discovery of the
  high $p_T$ electron events would be a firm prove of the neutralino
  DM component in halo.  In the experiment, the electron beam energy
  must be tuned within $O(10)$ MeV and the electron beam with high
  currents of $O(100)A$ is required for the detectors of the total
  length of a few hundred meters so that the sufficient event rate is
  obtained.  The dependence of the event rate on the DM velocity
  distribution in halo is also discussed.  This method might be
  applicable to other DM candidates.

\end{abstract}

\pacs{\bf 95.35.+d, 95.55.Ka, 11.30.Pb} 

\maketitle

Nature of the dark matter (DM) in the universe is an important problem
in particle physics, astronomy and cosmology. The lightest neutralino,
$\tilde{\chi}^0$, in the minimal supersymmetric (SUSY) extension of
the standard model (MSSM), is a good DM candidate since the lightest
SUSY particle (LSP) is stable due to the $R$ parity
\cite{DMCR}. The cosmological DM abundance, which is now precisely measured
by the WMAP \cite{wmap}, constrains properties of the neutralino and
the SUSY particle mass spectrum if the DM is generated in the hot
thermal bath in the early universe \cite{msugra}. 

The neutralino works well as the cold DM in the structure formation in
the universe. High resolution $N$-body simulations show that the cold
DM hypothesis explains well the large structure of our universe 
\cite{nbody}. On the other hand, the DM distributions in smaller
scales than $O(1)$ Mpc are still unresolved. The local DM abundance
and the DM velocity distribution on the neighborhood of the solar
system are also not well constrained from the rotation curve
measurements. The direct DM detection on the earth and the indirect
detection of anomalous cosmic rays produced by the DM annihilation may
give clues to the problems.

The conventional DM direct detection relies on nuclear recoil in
nuclei-DM elastic scattering. The sensitivities of the proposed
experiments cover a significant part of the MSSM parameter
space. However, to evaluate the local DM density and velocity
distribution from the counting rates, precise determination of the
hadronic matrix elements is necessary. The cross section also depends
on various SUSY parameters such as relatively suppressed Yukawa
coupling of strange quark (or $\tan\beta$), heavy Higgs mass and
Higgs-neutralino-neutralino coupling and so on. To this end, positive
signals in the direct detections may not necessarily prove whether the
DM is the neutralino since they merely measure the DM-nuclei scattering cross
section.

The nature of the LSP would be measured if the SUSY particles are
discovered in future high energy colliders.  This is because the LSP
will be copiously produced from the cascade decays of the heavier SUSY
particles. The LHC experiment is scheduled to start on 2007, and
squarks and gluino with masses up to 2.5~TeV can be discovered.
Furthermore, the interaction of the lightest neutralino would be also
measured at the LHC \cite{Nojiri:2005ph} and a future linear collider
(LC) \cite{Battaglia:2005ie}.  While they might be successful to
determine the thermal relic density of the universe and provide
consistency checks of the neutralino DM assumption, it relies on the
assumptions that the neutralino is stable in the cosmological time
scale and the thermal history of the universe follows the standard big
bang scenario.  It is important to observe the neutralino DM in more
direct and less ambiguous ways.

In this paper we investigate a possibility of direct neutralino DM
detection which might be realized in the future electron
accelerators. The electrons in the beam can interact with the DM
neutralinos in our neighborhood.  The electron-neutralino elastic
scattering is induced by the $s$-channel exchange of selectrons
$\tilde{e}^-$, which are superpartners for electron, in addition to
the $t$-channel $Z$ gauge boson exchange.  If the beam energy would be
tuned to the difference between the neutralino and selectron masses
($\Delta m$), the elastic scatterings between the DM neutralinos and
the beam electrons are dominated by on-pole selectron exchange and the
cross section is suppressed only by square of $\Delta m$. Building of
such an electron beam might be considered when the mass difference is
precisely measured in the future collider experiments such as LC
experiments.

If a high intensity electron beam would be available, one could
determine the nature of the DM without any ambiguities by detecting
on-pole production of selectrons, because the measurement proves that
the DM is the neutralino LSP.  Note that the relevant couplings would
be directly measured at LC when selectron productions are
accessible.  Therefore, the DM physics that might be explored by a high
intensity electron beam is unique. However, one needs several
technical breakthroughs  to realize it.  We clarify the requirements
to the electron beam and detectors in this paper. It is also shown
that the number of events might have a sensitivity to the DM
velocities when the selectron decay width is suppressed by the mass
difference or the coupling.  This implies that if there would be
enough statistics, the DM velocity distribution in the halo and the DM
wind due to a circular motion of the solar system might be constrained
by this experiment.

First, let us discuss the elastic scattering process between electrons
in the beam and the DM neutralinos and evaluate the expected number of
events.  The DM neutralinos are highly non-relativistic in the
universe, and the local DM velocity  in our neighborhood is typically $v\sim
10^{-3}c$. When the beam energy, $E_{\rm beam}$, would be tuned as $E_{\rm
beam}=\overline{E}_{\rm beam}$ with $\overline{E}_{\rm
beam}\equiv(m_{\tilde{e}^-}^2-m_{\tilde{\chi}^0}^2)/(2m_{\tilde{\chi}^0})(\simeq
\Delta m$),
the process is dominated by the on-pole selectron exchange.  In this
case the spin-averaged differential cross section with respect to the
angle between the beam and the recoiled electron, $\theta$, is given
as
\begin{eqnarray}
\frac{d\sigma }{d\cos \theta}
&=&
\frac{2\pi}
     {(m_{\tilde{e}^-}^2-m_{\tilde{\chi}^0}^2)^2}
 \frac{m_{\tilde{e}^-}^4}{m_{\tilde{\chi}^0}^2}
\frac{(m_{\tilde{e}^-}\Gamma_{\tilde{e}^-})^2}
     {(s-m_{\tilde{e}^-}^2)^2 + (m_{\tilde{e}^-}\Gamma_{\tilde{e}^-})^2}
\nonumber\\
&&\times {(1+A(\cos\theta))^{-2}}
\label{dsigma}
\end{eqnarray}
with $s$ square of the center-of-mass energy and $\Gamma_{\tilde{e}^-}$ the 
selectron decay width. Here, the function $A(\cos\theta)$ is defined as
\begin{eqnarray}
A(\cos\theta)
&=&\frac
{{m_{\tilde{e}^-}^2-m_{\tilde{\chi}^0}^2}}{{2 m_{\tilde{\chi}^0}^2}}
(1-\cos \theta),
\label{afunc}
\end{eqnarray}
and the energy of the recoiled electron, $E_{\rm recoil}$, is also given
by it as
\begin{eqnarray}
E_{\rm recoil}&=&E_{\rm
beam} (1+A(\cos\theta))^{-1}.
\end{eqnarray}
 When $\tilde{\chi}^0$ is bino-like, the
selectron decay width  is
\begin{eqnarray}
\Gamma_{\tilde{e}^-}&=&\frac{g_Y^2 Y^2}{8\pi}(O_{11})^2
\frac{(m_{\tilde{e}^-}^2-m_{\tilde{\chi}^0}^2)^2}{m_{\tilde{e}^-}^3}
\label{width}
\end{eqnarray}
where $Y$ is the hypercharge for $\tilde{e}^-$ and $O_{11}$ is the
neutralino mixing matrix element. For simplicity, we take $O_{11}=1$
and $Y=-1$ for the right-handed selectron in the following. In this case
\begin{eqnarray}
\Gamma_{\tilde{e}^-} 
&=&20~{\rm MeV}\times 
\left(\frac{\Delta m}{10{\rm GeV}}\right)^2
\left(\frac{m_{\tilde{e}^-}}{100{\rm GeV}}\right)^{-1}.
\label{gamma}
\end{eqnarray}

When the neutralino and selectron masses are close to each other, we
find from Eq.~(\ref{dsigma}) that the cross section at $E_{\rm
  beam}=\overline{E}_{\rm beam}(\simeq \Delta m)$ is suppressed only
by $(\Delta m)^2$. If the DM velocity dependence of the cross section
is negligible, the expected number of events, $N$, is
\begin{eqnarray}
N&=&
73\times
\left(\frac{\Delta m}{10{\rm GeV}}\right)^{-2}
\left(\frac{m_{\tilde{\chi}^0}}{100{\rm GeV}}\right)^{-1}
\left(\frac{\rho_{\rm DM}}{0.3{\rm GeV/cm^3}}\right)
\nonumber\\
&&
\times\left(\frac{j}{100 A}\right)
\left(\frac{T}{1~{\rm year}}\right)
\left(\frac{L}{1~{\rm km}}\right) .
\label{number}
\end{eqnarray}
Here, $j$ is the beam current, $L$ the detector length, and $T$ the
duration of experiment. We fix the local DM mass density, $\rho_{\rm
  DM}$, to be 0.3GeV/cm$^3$ in this paper. If the beam can be  polarized,
the expected number of events is multiplied by a factor two. This
elastic scattering process is also noticed in an earlier work in
Ref.~\cite{Aminneborg:1990xw}.

When $\Delta m/m_{\tilde{\chi}^0}\ll 1$, the elastic cross section can
be enhanced if the beam energy could be tuned to $\overline{E}_{\rm beam}$,
however, the requirements for the DM direct detection in electron
accelerators would be severe. We now  discuss the requirements to 
observe such events in order.

First, for our purpose, the neutralino and selectron masses and the
coupling have to be measured with sufficient precision at the earlier
collider experiments so that $\overline{E}_{\rm beam}$ and
$\Gamma_{\tilde{e}^-}$ are determined. Especially, the uncertainty of
$\overline{E}_{\rm beam}$ must be smaller than $\Gamma_{\tilde{e}^-}$.
Otherwise, the beam energy could not be tuned, so that the elastic
process between the DM neutralinos and the beam electrons is not
dominated by the on-pole selectron exchange. This implies from
Eq.~(\ref{gamma}) that $\overline{E}_{\rm beam}$ should be determined
at least with precision of $O(10^{-3})$ for $\Delta
m/m_{\tilde{\chi}^0}=10\%$.

At the LHC, the mass difference between the neutralino and sleptons
might be measured with the error on the order of a few GeV by using the
events with lepton-antilepton pair for favorable parameters
\cite{Bachacou:1999zb}. The LC would be able to measure the absolute
LSP and slepton masses with the error $O(50)$~MeV using the threshold
scan and the end point measurements \cite{Aguilar-Saavedra:2001rg}.
The error for the mass difference $\Delta m$, determined by the end
point measurement, may be even smaller. When the beam energy spread is
a dominant uncertainty in determination of $\Delta m$,
$\Delta m$ might be determined with the precision of $10^{-3}$ at the LC.
The neutralino interaction would be also measured precisely
at the LC so that the selectron decay width might be determined.

Next is the requirements for the electron beam.
The energy spread of the electron beam
must be less than $O(10^{-3})$ for $\Delta m/m_{\tilde{\chi}^0}=10\%$
by the same reason as for the determination of $\overline{E}_{\rm beam}$.
In addition, very high currents are required in our proposal
as seen in Eq.~(\ref{number}),
which is about 10 times higher than that for the
currently proposed Super $B$ factory.

In the KEKB at KEK, which is an asymmetric electron-positron collider
for $B$ physics, the averaged beam currents in the low energy positron
(3.5 GeV) and high energy electron rings (8.0 GeV) are 1.861$A$ and
1.275$A$, respectively.  The PEP-II at SLAC also has comparable
currents. Now upgrade of the KEKB to the SuperKEKB is proposed, aiming
for the beam currents 9.4 and 4.1$A$ \cite{Hashimoto:2004sm}. While
the energy spreads of these accelerators are smaller than $10^{-3}$, 
their beam currents are not much enough for our purpose.

The beam focusing at the interaction point is the sources of the beam
instability in collider experiments with high luminosities.  For the
DM direct detection in electron accelerators, it would not be  a serious
problem since the beam focusing is not required.  Rather, the
synchrotron radiation (SR) from the electron beams would bound the beam
currents. The SR at the arc sections of the accelerators may damage
the beam pipe and also causes the beam power loss. 

A hint to solve this problem may be in a technology called Energy
Recovery Linac (ERL) \cite{ERL}.  In this scheme, the electron beam energy
is lowered by transferring the energy to the RF power, and the power
is used to accelerating the electron beam again. The principle of the
ERL technology have been tested at Thomas Jefferson National
Accelerator Facility \cite{TJNA}.  Various facilities using the ERL
technology, such as photon factories, electron cooling and so on, are
proposed. If it would be possible to have a storage ring
where the beam energy is lowered before the arc section
to keep the beam power in the RF for accelerating
the electrons in the straight section,
one could achieve the high current electron beam with less power consumption
\footnote{
The scheme may be called as energy recovery storage ring
\cite{oide}. Naive estimation shows that the high frequency RF modes
may be induced to cause energy losses. 
}.

Even if an electron beam of $O(100)A$ is possible, the number of
events would be only around 1 event/10m/year for mass difference of
the order of 10 GeV. Note that because this is a fixed target
experiment where the target is on the beam line, many detector units
should be placed along the beam line. The total length must be as long
as 300 m$-$1 km. This means that each detector units should be as
simple as possible.

We note that the signal events should be distinguished over the
backgrounds (BGs). When $\Delta m/m_{\tilde{\chi}^0}\ll 1$, the
recoiled electrons in the signal events have energies between $E_{\rm
  beam}(1-2\Delta m/m_{\tilde{\chi}^0})$ and $E_{\rm beam}$ and large
transverse momentums. This is because $A(\cos\theta)$ in
Eq.~(\ref{afunc}) is suppressed by $\Delta m/m_{\tilde{\chi}^0}$ and
the angular distribution of the signal electron is almost
spherical. Furthermore, the momentums of the signal electron must be
pointing to the beam line.  The kinematics of the signal events is
well-constrained.

The signal electrons would be detected by placing either
electromagnetic calorimeter or tracking chamber with solenoid magnets
along the beam line. The expected BGs would come from either cosmic
rays or the beam interaction.  To remove accidental cosmic ray BGs, we
need a reasonable resolution for electron tracks to reject any tracks
crossing the beam line from outside of the detector.

Other BGs are electrons scattered by the beam gas.  They could be 
reduced by measuring the electron momentums since the BG electrons are
forward-going and $p_T\ll E_{\rm beam}$. The electron momentum
perpendicular to the beam axis can be measured precisely with magnetic
fields, while the resolution of the electron momentum along the beam
axis would be much worse than that in ordinary wire chambers.

Since the magnetic field along a long detector will destabilize the
beam, the detector should be divided to short segments along the beam
line so that the direction of the magnetic field in the each segment
is reversed to the next segment. This segmentation would also suppress
the BG electrons due to pile up from the upper currents, and the
background level might be lowered.

$\pi^-$s produced from photo-nucleon interactions in the beam pipe
would be also a source of the BGs. They are even more low energy compared
to the electrons from beam-beam gas interactions, and multiply
produced.  The detectors with $\pi^-$--$e^-$ separation abilities may
be useful to reduce the BGs furthermore.  Transition radiation
detectors (TRD) may have a benefit for the purpose. It is composed of
layers of radiators and X-ray detectors. The radiators emit soft X
rays when charged particles are injected. The radiation efficiency for
$\pi^-$ is lower than that for $e^-$ for $0.5$~GeV$ <p<100$~GeV, and
the information may be used to discriminate $\pi^-$s over $e^-$s.  The
rejection efficiency of about 99\% may be achieved for a total TRD
width $\sim$ 50cm \cite{PDG}.

We have sketched a rough design for the detector. While the signal
events have the well-defined kinematics and they might be
discriminated from the BGs by such a detector, the signal event rate
would be small and we need a long duration of the experiment. The
performance of the detector must be studied very carefully{, because 
the dominant part of the BGs must come from 
mis-measurements. }

We have neglected the dependence of the DM velocity distribution in
our neighborhood on the event rate in the above discussion.  However,
when the selectron decay width is small, the dependence cannot be
neglected.  When $E_{\rm beam}=\overline{E}_{\rm beam}$, the deviation
of $\sqrt{s}$ from $m_{\tilde{e}^-}$ in the signal process is
typically
\begin{eqnarray}
\sqrt{s}-m_{\tilde{e}^-}
&\sim& 10{\rm MeV} \times \left(\frac{\langle
v_{\|}\rangle}{10^{-3}}\right)\left(
\frac{\Delta m}{10{\rm GeV}}\right)
\label{dels}
\end{eqnarray}
with $\langle v_{\|}\rangle$ the average of DM velocity along the beam
axis.  It is found from Eq.~(\ref{width}) that this value is
comparable to or larger than the selectron decay width for $\Delta
m/m_{\tilde{\chi}^0}\lsim ~10\%$. Thus, the DM velocity distribution
affects the expected event rate when $\Delta m/m_{\tilde{\chi}^0}\lsim
~10\%$.

If the DM neutralinos are discovered in the electron accelerator
experiment, we might constrain the DM velocity distribution by
collecting the signal events. Since our DM direct detection in the
electron accelerator needs several breakthrough technologies, it might
be premature and speculative to discuss the measurement of the DM
velocity distribution in addition to the DM detection.  However, since
technologies to measure the DM velocity distribution in the
conventional DM detections are still limited, especially in cases of
the small counting rate \cite{Morgan:2004ys},  the
measurement of the DM velocity distribution using the electron
accelerator might be considered as one of the alternative 
possibilities.

In the following we evaluate the averaged cross section  assuming the
spherically symmetric isothermal sphere (SSIS) model for the DM
velocity distribution. The DM velocity distribution in the rest frame
of the earth is then
\begin{eqnarray}
f_h(\vec{v}) d^3 v
&=&
\left(\frac{3}{2\pi \sigma_h^2}\right)^{3/2}
e^{-\frac32 \frac{(\vec{v}+\vec{v}_{eh})^2}{\sigma_h^2}} d^3 v.
\label{dist}
\end{eqnarray}
We take the velocity dispersion of our local halo as $\sigma_h=270{\rm
km/s}$. $\vec{v}_{eh}$ is the velocity of the earth with respect to
the halo, and it is given as $\vec{v}_{eh}=
\vec{v}_{es}+\vec{v}_{sh}$, with $\vec{v}_{es}$ and $\vec{v}_{sh}$
the velocities of the earth with respect to the sun and of the sun
with respect to the halo, respectively. The solar system is moving toward the
constellation Cygnus ( $(\alpha,\delta)$=($21h$$12m$$01.053s$R.A.,
$+48^{\circ}$$19'$$46.71''$decl.) (J2000.0) in the equatorial
coordinate system) with speed $|\vec{v}_{sh}|=233 {\rm km/s}$.  The
earth's speed in the orbital motion is $|\vec{v}_{es}|=29.8 {\rm
km/s}$
\cite{Lewis:2003bv}. The motion of the solar system generates the DM
wind from the constellation Cygnus to the observers on the earth.
When the beam axis is parallel (perpendicular) to the DM wind,
$\langle v_{\|}\rangle$ is $\sqrt{\sigma_h^2/3+v_{eh}^2}$
($\sqrt{\sigma_h^2/3}$). 

Being the  experiments on the ground,
the beam axis 
rotates around the the earth
rotation axis. It leads to modulation of the event rate with a  period a
sidereal day ($23h56m4.09s$) due to the DM wind. Let us assume that
the azimuth angle from south for the beam axis is $\theta$ and the
experimental site is placed at latitude $\psi$. The angle between the
beam axis and the direction of the DM wind, $\Theta$, is modulated as
\begin{eqnarray}
\cos\Theta
&=&
\cos\delta_\star \cos\delta \cos(t-\alpha-t_\star)+ \sin\delta_\star
\sin\delta
\label{deltastar}
\end{eqnarray}
where $\sin\delta_\star = \cos\theta \cos\psi$ and $\sin t_\star
=-\sin\theta/\cos\delta_\star$, and $t$ is the local sidereal time at
the experimental site.

\begin{figure}[t]
\includegraphics[width=2.5in, angle=270]{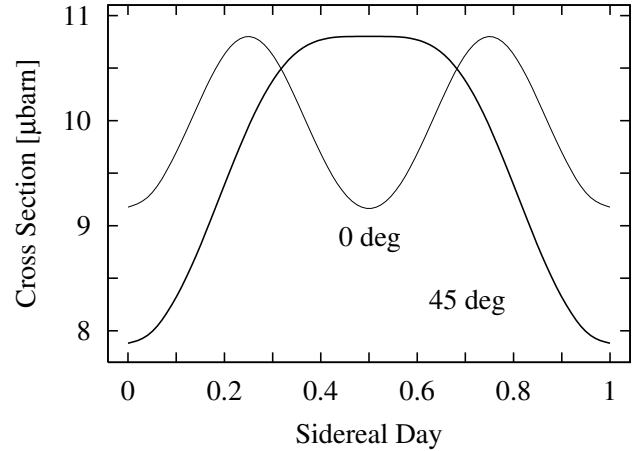}
\caption{
Modulation of cross section for $e^-\tilde{\chi}^0\rightarrow
e^-\tilde{\chi}^0$ during a sidereal day for $\delta_\star =0^{\circ}$ and
$45^{\circ}$. Here, we take $m_{\tilde{\chi}^0}=100$GeV, $\Delta
m=10$GeV and $E_{\rm beam}=\overline{E}_{\rm beam}$. Other
astrophysical parameters are given in text. }
\label{degeee}
\end{figure}

In Fig.~\ref{degeee}, the modulation of the cross section for the
signal event during a sidereal day is presented in cases of
$\delta_\star=0^{\circ}$ and $45^{\circ}$.  Here,
$m_{\tilde{\chi}^0}=100$GeV, $\Delta m=10$GeV, and $E_{\rm
  beam}=\overline{E}_{\rm beam}$. The cross section averaged by
Eq.~(\ref{dist}) is reduced from that for the on-pole selectron
exchange process ($\sim 12\mu {\rm barn}$). The latitude for the
experiment site $\psi$ is assumed to be $0^{\circ}$ for
simplicity. The phases for $\delta_\star=0^{\circ}$ and $45^{\circ}$
are not equal to each others when $\psi$ is different from
$0^{\circ}$. (See Eq.~(\ref{deltastar}).)  Since $\delta$ is close to
$\pi/4$, the amplitude of modulation in $\delta_\star=45^{\circ}$ is
almost maximum. The maximum (minimum) point on the curve corresponds
to a case that the beam is almost perpendicular (parallel) to the DM
wind. The angle $\alpha$ can be inferred by the phase of the
observed modulation.  It might be difficult to constrain $\delta$
using a single beam line with a fixed energy due to the parameter
degeneracy with $|\vec{v}_{eh}|/\sigma_h$. The degeneracy might be
resolved by using two beam lines with different $\delta_\star$s. It
is also possible to check the consistency of the observed DM wind
with the astrophysical observation.

\begin{figure}[t]
\includegraphics[width=2.5in, angle=270]{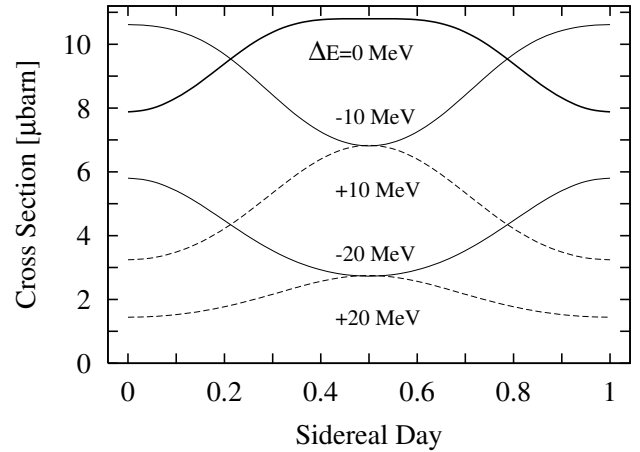}
\caption{
Modulation of cross section for $e^-\tilde{\chi}^0\rightarrow
e^-\tilde{\chi}^0$ during a sidereal day for the beam energy $E_{\rm
beam}=\overline{E}_{\rm beam} -20$, $-10$, $+0$, $+10$, and
$+20$MeV. Here we take $m_{\tilde{\chi}^0}=100$GeV, $\Delta m=10$GeV
and $\delta_\star =45^{\circ}$.  }
\label{energy}
\end{figure}

In the above, we assumed that the beam energy is tuned to
$\overline{E}_{\rm beam}$. The measurement of the selectron and
neutralino masses with precision 
of the order of or beyond an $O(10)$ MeV level in the
future collider experiments might be challenging, but important for
this experiment.  This can be seen in Fig.~\ref{energy}, where the
modulation of the signal cross section is presented in cases of
$E_{\rm beam}=\overline{E}_{\rm beam}-20$, $-10$, $+0$, $+10$, and
$+20$MeV. The cross section reduces significantly once $\vert \Delta
E\vert \gg20$~MeV for the parameters.  If the error of the mass
difference is more than 20 MeV, one has to scan the beam energy
to find the signals. The measurement of the beam energy dependence of
the event rate might be useful to determine the $\overline{E}_{\rm
  beam}$, since the phases are reverse in the positive and negative
energy deviation. Precise determination of $\overline{E}_{\rm beam}$
might allow us to interpret the event rate and determine the DM parameters.
Especially, the measurement in the different beam energies may be used
to resolve the parameter degeneracy between $\rho_{\rm DM}$ and
$\sigma_h$ in the observed event rate.

\begin{figure}[t]
\includegraphics[width=2.5in, angle=270]{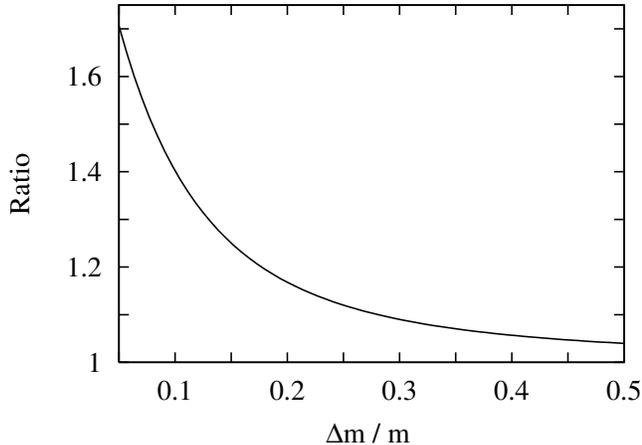}
\caption{
Ratio of cross sections for $e^-\tilde{\chi}^0\rightarrow
e^-\tilde{\chi}^0$ in two cases that the beam axis is parallel and
perpendicular to the DM wind, as a function of $\Delta m /m$. 
Here,  $E_{\rm beam}=\overline{E}_{\rm beam}$. }
\label{delmdep}
\end{figure}

The SSIS model has two parameters, $\rho_{\rm DM}$ and $\sigma_h$, in
addition to the velocity and direction of the DM wind. We have seen
that these might be constrained by changing the beam axis and/or beam
energy in the electron accelerator experiment, though a sizable event
number is required. We stress that it might be possible
when the selectron decay width is suppressed by the selectron
and neutralino mass degeneracy or the coupling
(or the neutralino mixing) and it is comparable to the
typical deviation of $\sqrt{s}$ from the selectron mass in the DM
velocity distribution.  In Fig.~\ref{delmdep} we show a ratio of the
cross sections in two cases that the beam axis is parallel and
perpendicular to the DM wind as a function of $\Delta m/m$.  Here, we
use Eq.~(\ref{gamma}) for the selectron decay width.  Larger $\Delta
m/m$ makes two cross sections closer since the selectron decay width
is increased. In this case, the modulation of the cross section
becomes featureless.

The large-scale features of the flat rotation curves around galaxies
are reproduced in the SSIS model.  However, numerous dynamical
arguments suggest that actual halo model may not be well described by
such a distribution. In addition to the axisymmetric \cite{axi} and
triaxial halo models \cite{triaxial}, non-Maxwellian distributions,
such as in the Sikivie caustic model \cite{sikivie}, are proposed.
The DM streams expected from the Sagittarius dwarf tidal stream might
also affect the local DM velocity distribution \cite{Morgan:2004ys}. It is
important to measure the local DM velocity distribution, including the
directional dependence, so that the models are discriminated. Our
proposal might be applicable to it.

In this paper we discuss a possibility of neutralino DM direct
detection through the scattering of local DM neutralinos with high
intensity electron beam.  The merit of the approach is that the DM can
be identified as "the lightest neutralino" in direct and less
ambiguous ways than the conventional DM detection experiments.
Furthermore, the local DM density and velocity distribution might be
constrained. However, to study the neutralino DM, the current of the
electron beam must be $O(10)$ times higher than those currently
planned, and the neutralino and selectron masses must be known very
precisely at the level of precision expected in the future linear
collider.  In addition, the mass difference between them should be
small to have a significant cross section.  The experiment also
requires the special setup to allow the long detector system along the
beam line.

\underline{Acknowledgment}\\
We would like to thank to Prof. K.~Oide for pointing out possibilities
to realize high intensity electron storage rings by using ERL. We
would also like to thank to N.~Toge, T.~Nakaya, M.~Hazumi, H.~Yamamoto
and T.~Yamanaka for various discussion, comments and encouragements.
This work is supported in part by the Grant-in-Aid for Science
Research, Ministry of Education, Science and Culture, Japan
(No.~13135207 and 14046225 for JH and No.~14540260, 14046210 and
16081207 for MMN).  MMN is also supported in part by a Grant-in-Aid
for the 21st Century COE "Center for Diversity and Universality in
Physics".

\end{document}